# Enhancing the efficiency of protein language models with minimal wet-lab data through few-shot learning

Ziyi Zhou[1†], Liang Zhang[1†], Yuanxi Yu[1†], Mingchen Li[1,2,3], Liang Hong[1,2,4]*, Pan Tan[1,2]*

1. Shanghai National Center for Applied Mathematics (SJTU Center), & Institute of Natural Sciences, & School of Physics and Astronomy, Shanghai Jiao Tong University, Shanghai 200240, China

2. Shanghai Artificial Intelligence Laboratory, Shanghai 200030, China

3. School of Information Science and Engineering, East China University of Science and Technology, Shanghai 200240, China

4. Zhang Jiang Institute for Advanced Study, Shanghai Jiao Tong University, Shanghai 201203, China

## Abstract

Accurately modeling the protein fitness landscapes holds great importance for protein engineering. Recently, due to their capacity and representation ability, pre-trained protein language models have achieved state-of-the-art performance in predicting protein fitness without experimental data. However, their predictions are limited in accuracy as well as interpretability. Furthermore, such deep learning models require abundant labeled training examples for performance improvements, posing a practical barrier. In this work, we introduce FSFP, a training strategy that can effectively optimize protein language models under extreme data scarcity. By combining the techniques of meta-transfer learning, learning to rank, and parameter-efficient fine-tuning, FSFP can significantly boost the performance of various protein language models using merely tens of labeled single-site mutants from the target protein. The experiments across 87 deep mutational scanning datasets underscore its superiority over both unsupervised and supervised approaches, revealing its potential in facilitating AI-guided protein design.

## Introduction

Proteins play an indispensable role in biological activities. Due to their attributes as biocatalysts, which are green, efficient, and cost-effective, the demand for their applications in scientific research and industrial production is steadily increasing[1-4]. However, most wild-type proteins, directly obtained from the biological species, cannot be directly applied in industrial conditions, as some of their physicochemical properties, such as stability, activity, and substrate specificity, are not good enough. Protein engineering seeks to excavate proteins with properties useful for specific applications. Traditional protein engineering, relying on methods like directed evolution and rational design, seeks to enhance these properties[5,6]. Directed evolution, although powerful, faces challenges in screening vast mutant libraries due to high-throughput assay constraints in terms of setup complexity and costs[7,8]. Rational design, despite its reduced experimental requirements, is often limited by the unavailability of detailed structural knowledge and mechanistic insights[9,10]. In recent years, deep learning has shown great potential in uncovering the implicit relationships between protein sequences and their functionality, i.e., fitness, thus being helpful to efficiently explore the vast design space.

Generally, deep learning approaches can be categorized into supervised and unsupervised models, with the main distinction being whether the training data requires experimentally collected labels[11-14]. Pre-trained protein language models (PLMs) are the most trending unsupervised approaches to fitness prediction. These models, such as ProGen[15], SaProt[16], ESM-2[14], and ESM-IF1[17], trained on the expansive protein universe, can estimate probability distributions for various protein sequences independent of experimental data. This capability facilitates the prediction of mutational effects but is limited in accuracy. Since these models fundamentally represent statistical characteristics of natural protein sequences found in nature, their zero-shot likelihood scores for mutation fitness essentially measure how similar a mutant protein sequence is to natural proteins or a particular protein family. While this measure can predict certain natural protein properties like solubility and stability[12,18,19], it inherently lacks the capability to predict non-natural catalytic properties, such as the catalysis of non-natural substrates or the production of non-natural products[20].

Supervised deep learning models, in contrast, have recently shown high accuracy in predicting protein fitness[11,13,21,22]. Based on their strong ability to extract both local and global features of the proteins, they could construct more accurate sequence-fitness correlations by training on sufficient labeled data. However, these models are heavily reliant on extensive data derived from expensive, high-throughput mutagenesis experiments[23,24], posing a significant challenge for most proteins. Recently, Hsu et al.[25] developed an efficient ridge regression model that combines the one-hot features of amino acids and the probability density feature calculated by an unsupervised model. It demonstrates improved performance against more sophisticated and expensive methods with limited experimental data. However, one-hot features are not informative enough to represent the relationships between different residues. Besides, as a linear model, ridge regression might have difficulty in learning complex patterns that affect protein fitness. Therefore, it is meaningful to develop new strategies to effectively fine-tune PLMs with scant wet lab data in protein engineering, where the advantages of both unsupervised and supervised approaches can be fused.

In this work, we leverage the synergistic methodologies of meta-transfer learning (MTL)[26], learning to rank (LTR)[27,28], and parameter-efficient fine-tuning[29,30] to develop a versatile approach for training PLMs. Our approach, named FSFP (**F**ew-**S**hot Learning for Protein **F**itness **P**rediction), is notable for its reliance on a minimal labeled dataset for the target protein, comprising merely tens of random experimental single-site mutants. This streamlined dataset is sufficient to substantially enhance the accuracy of the trained model to predict mutational effects. To validate our approach, we conduct simulation experiments using various representative PLMs, including ESM-1v, ESM-2, and SaProt. Although FSFP is theoretically compatible with any PLMs, our selection of models for testing is primarily influenced by practical considerations, particularly computational efficiency and resource constraints. Our methodology demonstrates remarkable performance on ProteinGym[31], a benchmark including an extensive range of 87 deep mutational scanning (DMS) datasets, showing robustness when adapting to different PLMs and proteins. This is achieved in comparison with both unsupervised and supervised models trained on tens of data. In particular, our approach can significantly enhance the models that initially have poor zero-shot performance for predicting the fitness of certain proteins (e.g., with Spearman correlations below 0.1), elevating their accuracy to an acceptable level (e.g., with Spearman correlations exceeding 0.5) even with as few as 20 experimentally labeled examples of the target protein. These enhancements underscore the robust transferability of FSFP across various proteins and its exceptional efficiency in data utilization. Such findings highlight the wide-ranging applicability and potential impact of FSFP within the realm of AI-guided protein engineering.

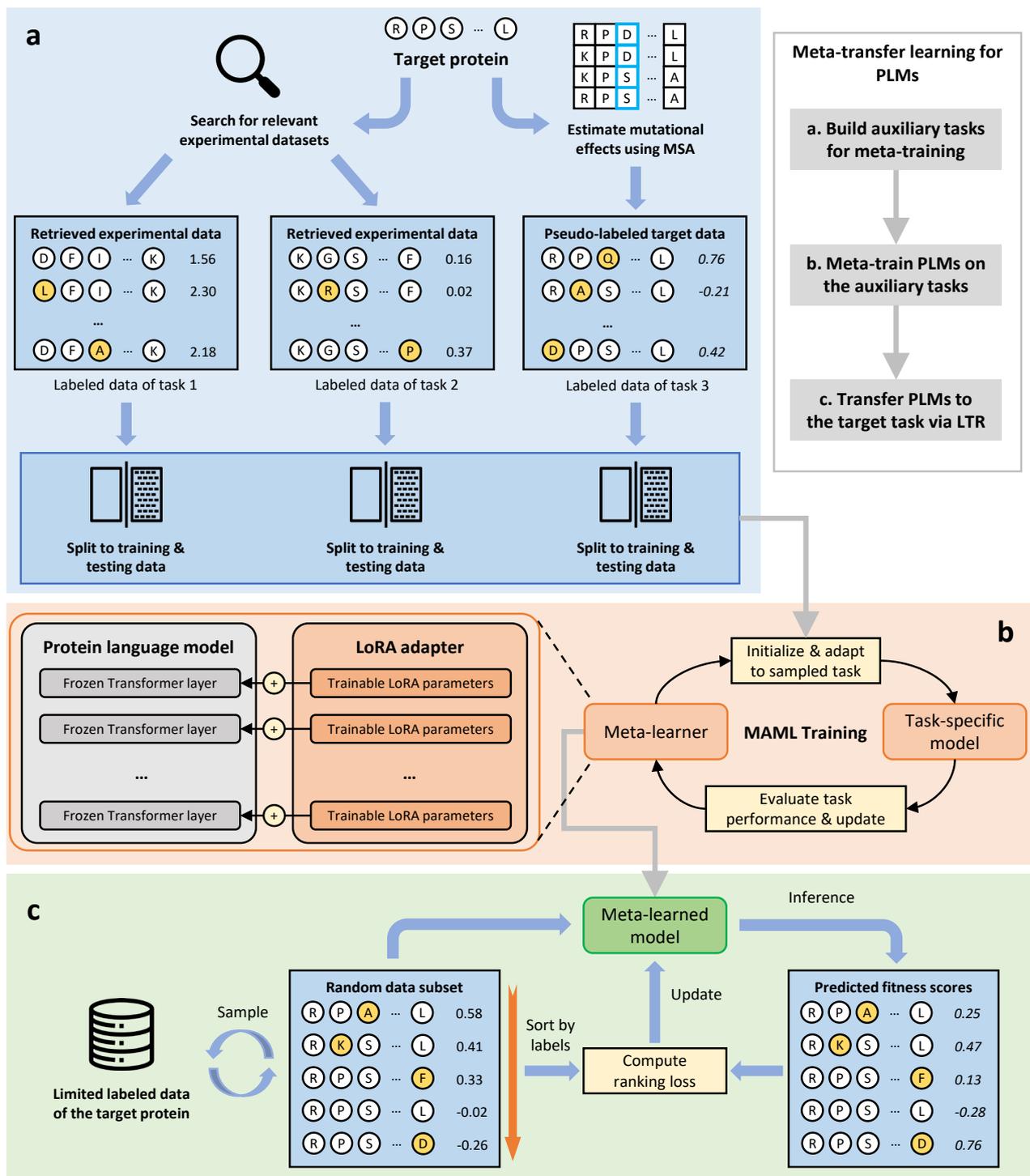

**Fig.1 Overview of FSFP.** FSFP includes three stages: building auxiliary tasks for meta-learning, meta-training PLMs on auxiliary tasks, and transferring PLMs to the target task. **a)** Based on the wild-type sequence or structure of the target protein, the labeled mutant datasets of two similar proteins are retrieved to be the first two tasks. In addition, an MSA-based method is used to estimate the variant effects of the candidate mutants as pseudo labels for the third task. **b)** MAML algorithm is used to meta-train PLMs on the built tasks, and eventually optimizes them into a meta-learner that provides good parameter initialization for the target task (right). To prevent PLMs from overfitting on small training data, LoRA is applied to constrain model updates to a limited number of parameters (left). **c)** The meta-trained model is then transferred to the target few-shot learning task. FSFP treats fitness prediction as a ranking problem, and leverages LTR technique for both transfer learning and meta-training. It trains PLMs to rank the fitness by computing a listwise ranking loss between their predictions and the ground truth permutation.

# Results

**Transferring PLMs with few-shot training data via FSFP**

FSFP leverages meta-learning to better train PLMs in a label-scarce scenario (Fig.1 and Methods). Meta-learning aims to train a model that can rapidly adapt to a new task using only a few training examples and iterations, by accumulating experience from learning multiple tasks[26,32,33]. To build the training tasks required for meta-learning, we search for existing labeled mutant datasets that are potentially helpful to predict the variant effects on the target protein, as well as generate pseudo labels through multiple sequence alignment (MSA) for the candidate mutants (Fig.1a). In this stage, the wild type sequences or structures of the target protein and the ones in the database are first encoded into embedding vectors by PLMs. ProteinGym is used as the database to retrieve due to the diversity and quantity of its mutant datasets. After that, the associated datasets of the top two proteins that are closest to the target protein in the vector space are selected to form the first two tasks. This is motivated by the fact that the fitness landscapes of similar proteins may share similar properties[34]. On the other hand, existing literature shows that it is promising to predict the effect of genetic variation using MSA[11,19,31,35-38]. We thereby use an alignment-based method, GEMME[35], to utilize MSA information of the target protein, and score the candidate mutants of interest to build the dataset of the third task. The labeled data of these tasks are randomly split into training and testing data respectively. In this way, we expect the meta-trained model can learn to utilize the target training data from both evolutionary information and similar fitness landscapes.

We apply model-agnostic meta-learning (MAML)[33] to meta-train PLMs on the built tasks (Fig.1b). In effect, MAML learns to find the optimal initial model parameters such that small changes in them will produce large improvements on the target task. The meta-training procedure has two levels of optimization in each iteration, and will eventually turn PLMs into a meta-learner for initialization (Fig.1b, right). In the inner-level optimization, a temporary base-learner is initialized by the current meta-learner, and then updated into a task-specific model using the training data of a sampled task. In the outer-level optimization, the test loss of the task-specific model on that task is used to optimize the meta-learner.

PLMs typically use heavily parameterized Transformer[39] models as the base architecture and pre-train on large-scale unlabeled protein sequences[12,14,15,19,40,41] or structures[16,17]. However, when fine-tuning on very few labeled training data, they are likely to suffer from catastrophic overfitting. Therefore, FSFP utilizes low-rank adaptation (LoRA)[29] to inject trainable rank decomposition matrices into PLMs with their original pre-trained parameters frozen, and all of the model updates are constrained to these small number of trainable parameters (Fig.1b, left).

After meta-training, a good initialization of the LoRA parameters is obtained, and we finally transfer the meta-trained PLMs to the target few-shot learning task, i.e., learning to predict the mutational effects of the target protein using its limited experimental data. Unlike conventional approaches to train supervised protein fitness predictors, which formalize this to a regression problem[11,13,21,22,25], FSFP treats it as a ranking problem and leverages the LTR technique (Fig.1c). In protein engineering and directed evolution, the most important index is whether a mutation enhances the functional fitness of the existing protein. Therefore, instead of focusing on the specific score values of mutations, their relative effectiveness or ranking order should hold greater significance. Specifically, FSFP learns to rank the fitness by computing ListMLE loss[28], which is defined by the likelihood of the permutation of a correct ranking. In each iteration, the model is trained to fix its predictions for one or more sampled data subsets to make them conform to the ground truth permutation. The above training scheme is adopted in both the transfer learning stage (using the target training data) and the inner-level optimization during the meta-training stage (using the training data of the auxiliary tasks).

**Experimental setup**

We evaluate model performance on the substitution benchmark of ProteinGym, which consists of about 1.5M missense variants from 87 DMS assays. Among these 87 datasets, 11 contain multi-site mutants. For each dataset in the benchmark, we first randomly select 20 single-site mutants as an initial training set. Then we expand the training set size to 40 by sampling another 20 single-site mutants. The training sets with sizes of 60, 80, and 100 are built accordingly. For each training set, all the remaining data (or part of them if specified) are used as test set, and we use cross-validation on the training data to determine the training hyperparameters (Methods). The data splitting process is repeated 5 times with different random seeds, and we average the model performance over different splits of a certain training size. The predictive performance is measured by two metrics: Spearman rank correlation and normalized discounted cumulative gain (NDCG)[42] with the experimental fitness labels as ground truth.

Our main baseline on few-shot protein fitness prediction is the ridge regression approach introduced by Hsu et al.[25]. It uses one-hot encoded site-specific residue features and the fitness score predicted by an existing evolutionary probability density model as the input features to train a ridge regression model. Although it is simple, it turns out to be more effective than other supervised learning approaches in such low-resource scenarios. When applying FSFP to a foundation model for evaluation, we compare it to the ridge regression-augmented version of this model. We use the official implementation of this baseline to ensure its performance.

**All components of FSFP contribute positively to few-shot learning**

To thoroughly evaluate the impact of different components that make up FSFP, we conduct an ablation study taking ESM-2 as the foundation model (Fig.2). In detail, we compare FSFP with the following training strategies. LTR + LoRA + MTL (no MSA) is a variant of FSFP that does not depend on MSA to build auxiliary tasks. It replaces the third task of FSFP with another labeled dataset retrieved from the database. LTR + LoRA is a variant of FSFP that utilizes the LTR technique only. It transfers the LoRA-adapted model to the target task without meta-training on auxiliary tasks. MSE denotes fine-tuning the entire PLM with fitness labels as done by Rives et al.[41], which uses the likelihood difference between the mutant and the wild-type as a predictor. MSE + LoRA further enhances the MSE method with LoRA. Ridge regression denotes the approach proposed by Hsu et al.[25]. We also perform zero-shot inference using the PLM following Meier et al.[12].

As illustrated in Fig.2, compared with zero-shot inference, MSE exhibits major degradation on the average predictive performance. This can be attributed to the quick overfitting caused by directly fine-tuning the whole model on small training datasets (Fig.S1b and Fig.S1c). Equipping LoRA significantly mitigates such negative effects overall due to its efficient parameterization for minimal adjustments. However, its performance still fails to match the original unsupervised model until the training set size reaches 80. As our main baseline, the performance of ridge regression exceeds zero-shot inference across most of the training set sizes. This is consistent with the comparison results presented by Hsu et al.[25]. Meanwhile, LTR + LoRA outperforms the above methods on all training set sizes in terms of both Spearman correlation and NDCG. As mentioned before, the ranking order of different mutants is often more important than their absolute scores in directed evolutions. The combination of LTR and LoRA is able to solve this task much easier than using regression since the model is not forced to fit the exact label values.

Using MTL to obtain initial model parameters for few-shot training further increases the performance. The improvements of LTR + LoRA + MTL (no MSA) over LTR + LoRA indicate that similar properties in fitness landscapes from other proteins can be helpful. FSFP, which additionally learns the MSA knowledge during meta-training, achieves the best scores on all training data sizes in terms of both metrics. This suggests that apart from the training data of the target protein, the evolutionary information from MSA may also effectively supervise the model in estimating mutational effects. As illustrated by the training curves (Fig.S1), with or without MSA, MTL could significantly improve the model performance using very few training iterations compared with other approaches. Besides, the initial model after meta-training can already substantially outperform zero-shot inference without access to the target training data (i.e., with no training iteration) in some cases. These demonstrate that by utilizing FSFP, the model successfully learns useful information from auxiliary tasks and thus can well transfer to the target few-shot learning task.

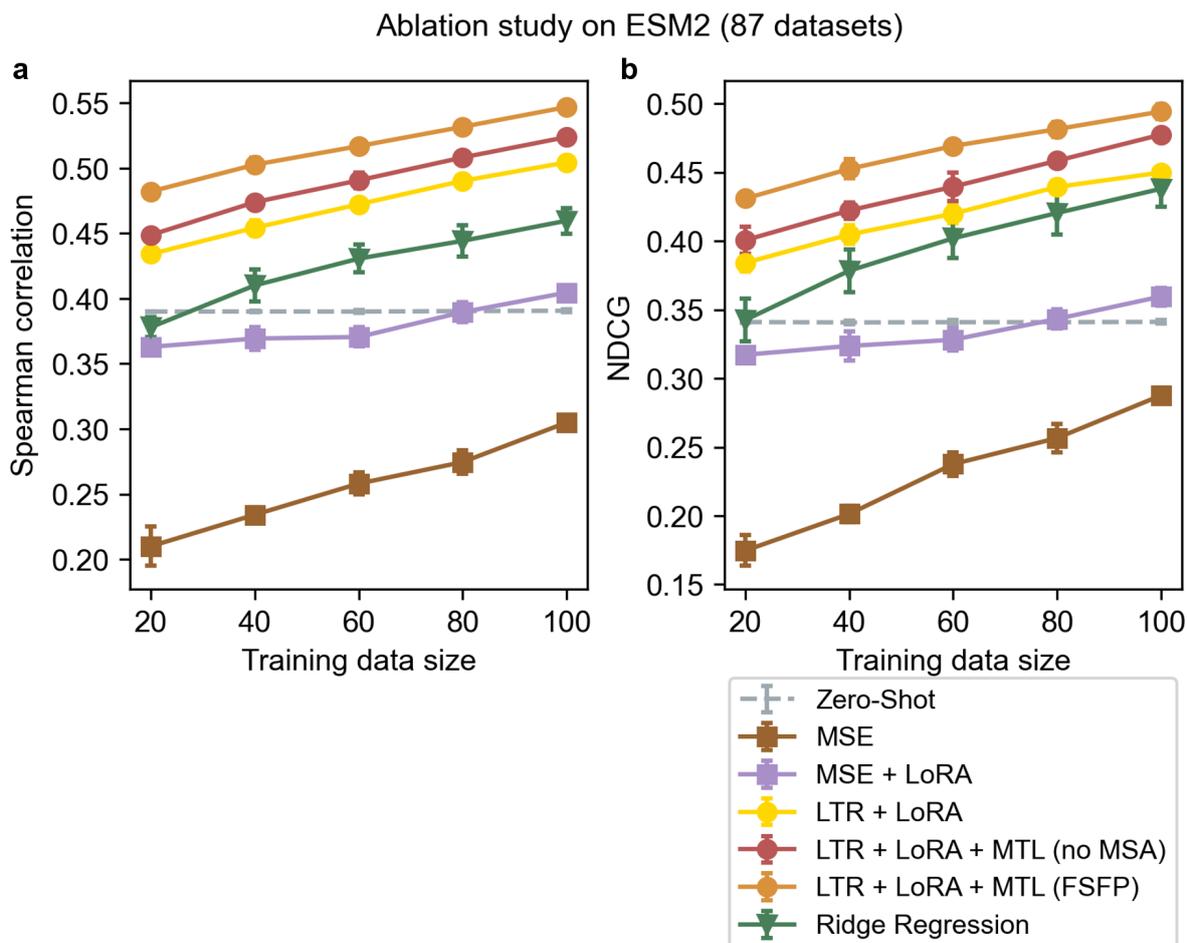

**Fig.2 Ablation study on ESM-2.** Average performance of each strategy across all datasets in ProteinGym with respect to the training data size, evaluated by **a)** Spearman correlation and **b)** NDCG. For each dataset, we randomly pick 20, 40, 60, 80, and 100 single-site mutants as training set, with all the rest as test set. Each dot in these figures is the average test performance of 5 random data splits, and the error bars represent the standard deviation caused by different splits. Two-sided Mann-Whitney $U$-tests are used for comparing the performance of FSFP with all other strategies, getting $P < 0.0079$ for all training sizes.

**FSFP as a general few-shot learning approach for PLMs**

FSFP can be applied to any deep learning-based protein fitness predictor that uses gradient descent for optimization, while we focus on PLMs in this work. To validate its versatility, we select three representative PLMs—ESM-1v, ESM-2, and SaProt—as the foundation models to be trained (Methods). For each of these models, we compare FSFP with their zero-shot predictions, as well as the ridge regression approach across all 87 datasets in the benchmark. Since FSFP leverages GEMME to generate pseudo labels for meta-training, we also add GEMME and GEMME augmented by the ridge regression approach as two additional baselines. The test performance on single-site mutants and multi-site mutants are reported separately (Fig.3 and Fig.S2). Note that the best approach reported by Hsu et al.[25] is DeepSequence[37] augmented by ridge regression. However, we find that its overall performance is obviously worse than the ridge regression-augmented GEMME on ProteinGym, so we omit this approach from the figures.

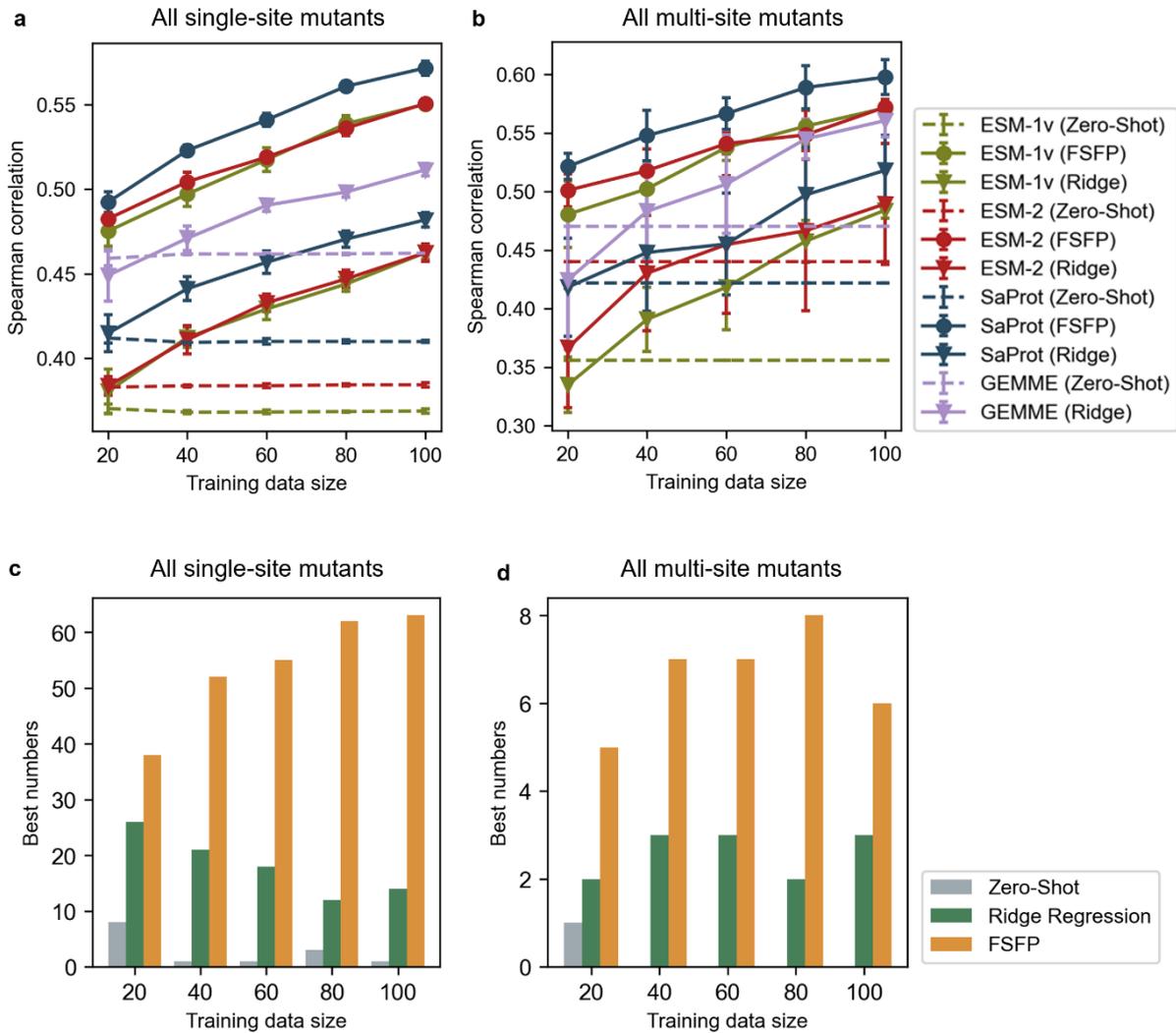

**Fig.3 Overall performance on single-site and multi-site mutants. a)** Average model performance tested on single-site mutants across all 87 datasets, evaluated by Spearman correlation. SaProt (FSFP) is significantly better than all baselines with $P < 0.0079$ (Mann-Whitney $U$-test) for all training sizes. **b)** Average model performance tested on multi-site mutants across 11 datasets, evaluated by Spearman correlation. SaProt (FSFP) is significantly better than all baselines with $P < 0.016$ (Mann-Whitney $U$-test) for all training sizes. Analogous results measured by NDCG are shown in Fig.S2a and Fig.S2b. **c)** Summary of how often the best test Spearman correlation for single-site mutants on a certain dataset is achieved by a PLM, where the colors represent different strategies applied to the best PLMs. **d)** Similar to Fig.3c, but counted for the best performance on multi-site mutants.

Considering the average performance, the PLMs trained by FSFP consistently outperform other baselines on all training data sizes (Fig.3a and Fig.3b). Among them, SaProt (FSFP) emerges as the top performer, while ESM-1v (FSFP) and ESM-2 (FSFP) show comparable performance. Besides, on most datasets in ProteinGym, the best Spearman correlation is achieved by one of the FSFP-trained PLMs (Fig.3c and Fig.3d). Compared with zero-shot predictions, FSFP boosts the performance of PLMs on single-site mutants by nearly 0.1 on Spearman correlation using only 20 training examples, and this gap becomes even larger when it comes to multi-site mutants. The improvements keep increasing as the training dataset expands, aligning with the results of the prior ablation study. This demonstrates the adaptability and effectiveness of FSFP on different foundation models. By contrast, the ridge regression approach fails to noticeably outperform its zero-shot counterparts with 20 training examples. On multi-site mutants, it shows major negative impacts on the performance of GEMME, ESM-1v, and ESM-2 when the training size is 20. Although its performance also tends to improve as the training size increases, it is consistently behind the FSFP-trained models. Except for the training strategy, this is likely due to the limitation of simple one-hot features and model capacity of the ridge regression approach. It is worth noting that when predicting multi-site mutants, the standard errors of FSFP are much smaller than those of ridge regression, indicating that the former is more stable and reliable when given different few-shot training data.

Notably, the average zero-shot performance of GEMME is better than the PLMs we choose. This is not surprising because PLMs may not always be better than MSA-based methods on predicting variant effects[25,31], despite their strong ability to model the statistical characteristics of protein sequences or structures. Therefore, resorting to guidance from a state-of-the-art MSA-based method is reasonable in such a few-shot scenario. As shown in Fig.3a, the models utilizing FSFP achieve substantial improvements over both GEMME and its ridge regression-augmented version on all training sizes. This suggests that FSFP not only instills the MSA knowledge from GEMME into PLMs, but also successfully combines it with the supervised information from the target training data through MTL. This again verifies the superiority of FSFP as a few-shot learning strategy, especially with exceedingly small training datasets.

**FSFP holds robust generalizability and extrapolation ability**

In practice, the effect of mutations whose positions do not occur in existing experimentally labeled data is always wondered. Therefore, the capability of the fitness predictors to extrapolate across positions is important for probing mutants with good properties from the enormous sequence space. We examine the extrapolation ability of different approaches by evaluating them on specific subsets of the testing data. In detail, from each original test set, we first select all single-site mutants whose mutated positions are different from those of the training examples, resulting in a more difficult test set of single-site mutants. Then we select the multi-site mutants whose individual mutations have no overlap with the mutations in the training data, resulting in another challenging test set. We do not force all mutated positions of a testing multi-site mutant to be different from the training positions because this will lead to insufficient testing examples for some datasets. Under these settings, we can find that the zero-shot performance of the base models obviously varies with the training set size (Fig.4). This can be attributed to the greater changes in the test sets: as the training dataset expands, available testing examples satisfying above conditions may drastically decrease for some datasets, impacting the evaluation results of zero-shot predictions.

When extrapolating to single-site mutants with different positions, the models augmented by ridge regression do not show clear improvements over the base models even with 100 training examples (Fig.4a). For multi-site mutants, the ridge regression approach fails to effectively enhance GEMME and ESM-2 when the training size is less than 60, and it exhibits larger standard deviations than FSFP as observed before (Fig.4b). In stark contrast, the PLMs trained by FSFP continue to score better than all baselines on Spearman correlation across different training sizes. Compared with their zero-shot performance, they show considerable improvements after being trained by FSFP, especially for ESM-1v. Also, the best performer of most datasets is an FSFP-trained model (Fig.4c and Fig.4d). These are consistent with the quantitative conclusions illustrated by Fig.3. Since PLMs have intrinsic strong generalizability owing to their high capacity and embedded knowledge through pre-training, it is possible to boost their performance on hard downstream tasks by proper training paradigm such as FSFP. However, the input features of the ridge regression approach are much less informative, resulting in limited extrapolation ability.

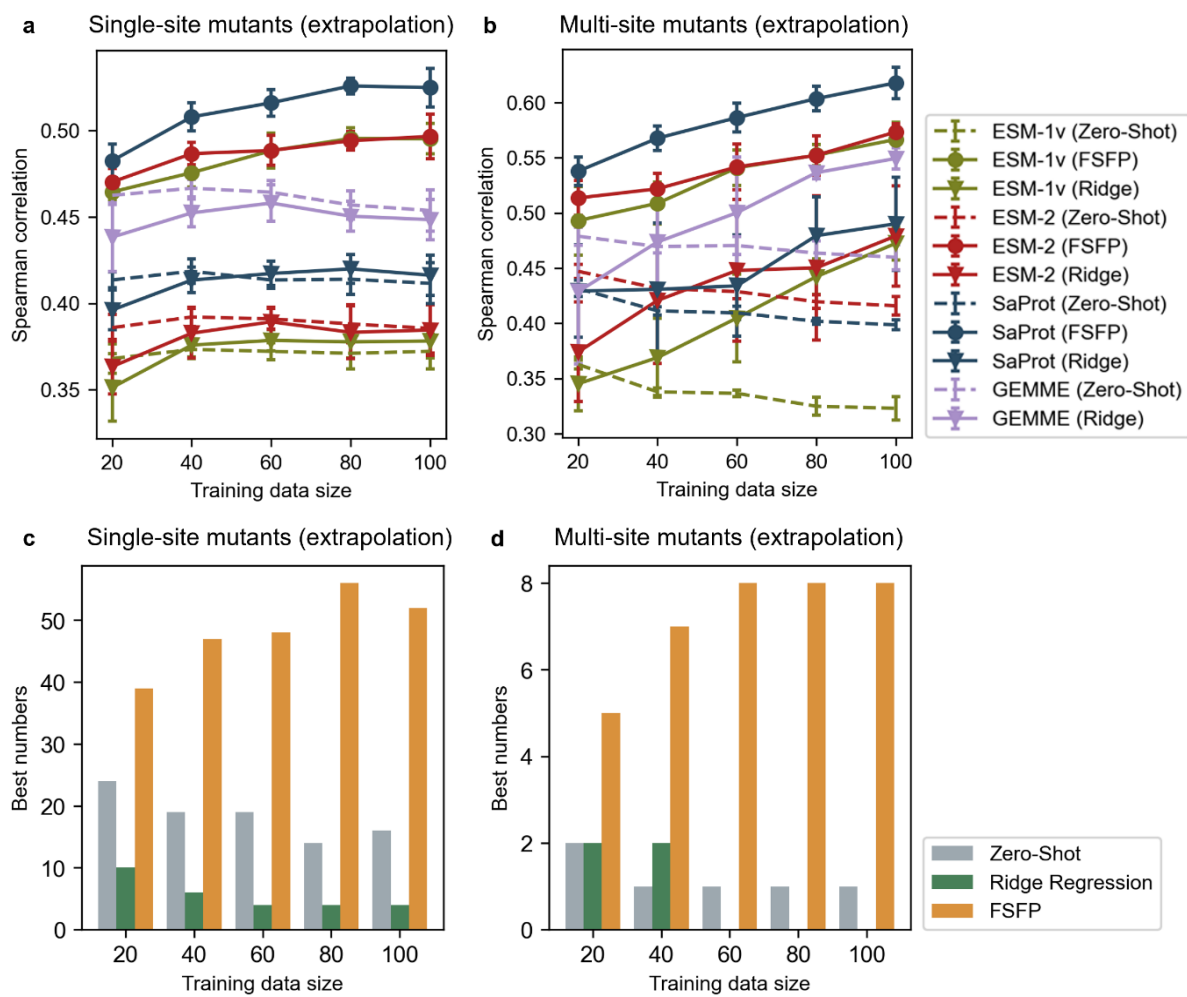

**Fig.4 Extrapolative performance on single-site and multi-site mutants. a)** Extrapolating to single-site mutants whose mutated positions do not occurred in the training set, evaluated by Spearman correlation. SaProt (FSFP) is significantly better than all baselines with $P < 0.016$ (Mann-Whitney $U$-test) for all training sizes. **b)** Extrapolating to multi-site mutants whose individual mutations have no overlap with the mutations in the training data, evaluated by Spearman correlation. SaProt (FSFP) is significantly better than all baselines with $P < 0.0079$ (Mann-Whitney $U$-test) for all training sizes. Analogous results measured by NDCG are shown in Fig.S3a and Fig.S3b. **c)** Summary of how often the best extrapolative Spearman correlation for single-site mutants on a certain dataset is achieved by a PLM, where the colors represent different strategies applied to the best PLMs. **d)** Similar to Fig.4c, but counted for the best extrapolative performance on multi-site mutants.

To further demonstrate the applicability and generalizability of FSFP, we show the comparison results of different approaches on four proteins: the envelope protein Env from HIV[43], the human α-synuclein[44], protein G (GB1)[45], and the human TAR DNA-binding protein 43 (TDP-43) [46]. In these cases, one or more unsupervised models show poor performance and are not reliable in practice, highlighting the need for effective training using limited experimental labels (Fig.5). Notably, for TDP-43, the zero-shot results of all models have no or even negative correlation with the actual fitness labels (Fig.5d), and those of ESM-2 have low accuracy for HIV Env (Fig.5a) and α-synuclein (Fig.5b). As seen, except for GB1 (Fig.5c), most models enhanced by ridge regression do not show significant improvements against their zero-shot performance, even with larger training datasets. On the contrary, utilizing FSFP, the PLMs achieve considerable gain by training on small datasets, therefore becoming more useful for directed evolution. It can also be observed that the performance of GEMME, the method used by FSFP for yielding pseudo labels, is not dominating among other base models in the later three examples. Nevertheless, this does not hinder the FSFP-trained models from being the top performers, which suggests that FSFP learns to generalize from the auxiliary tasks instead of simply overfitting them.

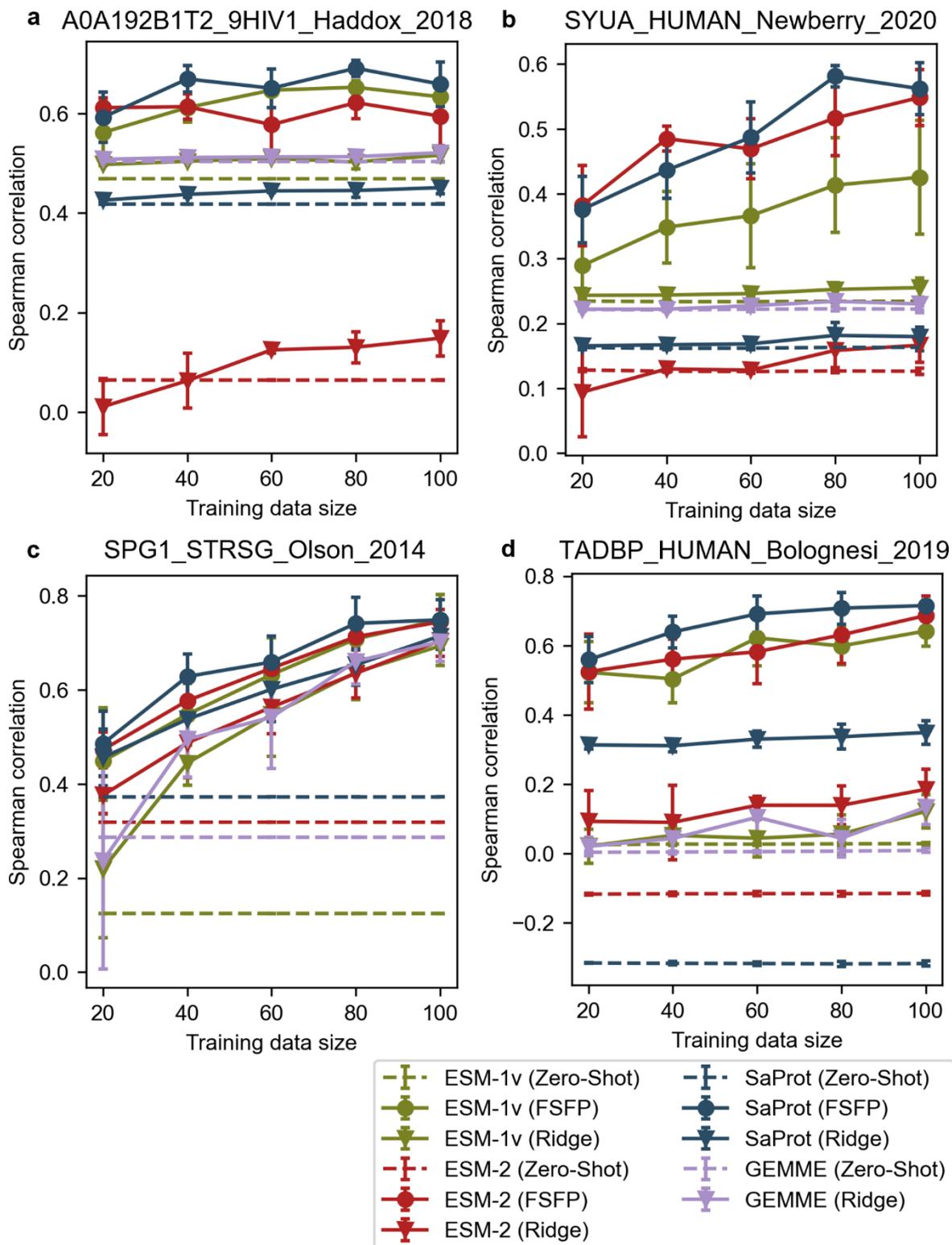

**Fig.5 Comparison of different approaches on four proteins. a)** The envelope protein Env from HIV. **b)** The human α-synuclein. **c)** Protein G (GB1). **d)** The human TDP-43. The models are trained on single-site mutants and tested on all remaining data using Spearman correlation. Analogous results measured by NDCG are shown in Fig.S4.

## Discussion

In this work, we introduce a new paradigm for effectively training PLMs to predict protein fitness using only a small number (tens) of experimentally labeled mutants. FSFP integrates the techniques of LTR, LoRA, and MTL, where LTR meets the intrinsic needs of directed evolutions to rank the protein fitness, LoRA greatly reduces the overfitting risk of PLMs when encountering small training datasets, and MTL provides PLMs with better initial parameters for fast adaptation to the target task. We apply FSFP to three representative PLMs, i.e., ESM-1v, ESM-2 and SaProt for the case study, though it is theoretically compatible with any PLM. Through comprehensive experiments across 87 DMS datasets, we demonstrate the effectiveness and robustness of FSFP in few-shot protein fitness prediction: 1) it boosts the test performance of the PLMs by up to 0.1 on average Spearman correlation using merely 20 training examples; 2) it improves the performance of different PLMs consistently and considerably; 3) it enables PLMs to well extrapolate to the mutations whose positions are absent in the training data; 4) it remains both effective and data-efficient even when the PLMs exhibit poor zero-shot performance on the target protein.

Compared with other foundation models trained by FSFP, SaProt shows the best performance in most cases. Since SaProt encodes both structural information and amino acid sequence of the target protein, it could capture more features that affect the protein function than its counterparts. However, it is noteworthy that no model can keep dominating others on all datasets after training, i.e., the best performer varies for different datasets. This suggests that different models suit differently when learning the fitness landscapes of various proteins. In practical applications, the choice of the foundation model to be trained should be given careful consideration. For example, when predicting the properties highly related to protein structure such as binding and thermostability, one can select SaProt for training.

As a preliminary work to leverage meta-learning for protein fitness prediction, we use a simple method, i.e., PLM-guided protein search, to find useful mutant datasets to build auxiliary tasks for meta-training. Intuitively, the improvements caused by FSFP could be more significant if the selected datasets are more relevant to the target dataset in both their wild-types and labeled properties. However, the similarities between the proteins in ProteinGym are low overall, and we thereby only select the top-2 datasets for meta-training to avoid negative transfer (we tried to include more datasets but the final performance did not noticeably increase). We argue that it is important to collect auxiliary datasets that are close enough to the target task based on prior knowledge if available. For example, using different types of experimental data from the target protein, or using the labeled data from its protein family.

Based on the superior few-shot performance, FSFP could enable more effective directed evolution, especially when high-throughput screens are difficult. The initial data for directed evolution may originate from rational design, random mutations, or zero-shot predictions from PLMs. Regardless of the proportion of positive mutants in this initial dataset, it could serve as a basis for selecting the most suitable PLMs. In subsequent iterative rounds, FSFP could be applied to train the selected PLMs. Leveraging the extrapolative capabilities of the models trained by FSFP, they can be applied to recommend new mutants.

## Method

**Efficient solution for the fitness ranking problem**

**Parameter-efficient finetuning of PLMs.** To prevent PLMs from overfitting on small training datasets, we use LoRA[29] to learn a small number of task-specific parameters instead of optimizing the full model. LoRA hypothesizes that the change of weights during model tuning has a low intrinsic rank. In detail, given a pre-trained weight matrix $\boldsymbol{W}_0 \in \mathbb{R}^{d \times k}$, LoRA constrains its update by representing the latter with a low-rank decomposition:

$$\boldsymbol{W}_0 + \Delta \boldsymbol{W} = \boldsymbol{W}_0 + \boldsymbol{B}\boldsymbol{A} \qquad (1)$$

where $\boldsymbol{B} \in \mathbb{R}^{d \times r}$, $\boldsymbol{A} \in \mathbb{R}^{r \times k}$, and the rank $r \ll \min(d, k)$. $\boldsymbol{A}$ is randomly initialized but $\boldsymbol{B}$ is initialized by zero, so $\Delta \boldsymbol{W}$ is also zero at the beginning of training. We freeze the original pre-trained model and apply LoRA to the weight matrices in its self-attention modules and feed-forward layers. The hyperparameter $r$ is set to 16 in all our experiments.

**Learning to rank the fitness.** We use a listwise LTR approach, namely ListMLE[28] to train PLMs on few-shot training data. It defines the loss function based on the likelihood of the permutation of a correct ranking. Let $\boldsymbol{x} = \{x_1, \ldots, x_n\}$ be the objects to be ranked, $\boldsymbol{y} = \{y_1, \ldots, y_n\}$ be their corresponding labels, and $\pi$ be a permutation of $\boldsymbol{x}$ satisfying $\pi \in \{\pi | y_{\pi(i)} \geq y_{\pi(j)}; i < j\}$ where $\pi(i)$ denotes the index of the object ranked at position $i$ in $\pi$. The ListMLE loss is defined as:

$$\mathcal{L}(f; \boldsymbol{x}, \pi) = -\log P(\pi | \boldsymbol{x}; f) \qquad (2)$$

where $f$ is a parameterized scoring function, and the probability $P(\pi | \boldsymbol{x}; f)$ is defined by the Plackett-Luce model[47]:

$$P(\pi | \boldsymbol{x}; f) = \prod_{i=1}^{n} \frac{\exp\left(f(x_{\pi(i)})\right)}{\sum_{k=i}^{n} \exp\left(f(x_{\pi(k)})\right)} \qquad (3)$$

Given a mutant $\boldsymbol{s}$, we score its fitness by comparing the probability assigned to each mutated residue by PLMs with the one for the wild type, following Meier et al.[12]:

$$f(\boldsymbol{s}) = \sum_{t \in T} [\log P(\boldsymbol{s}_t = \boldsymbol{s}_t^{mt} | \boldsymbol{s}^{mt}) - \log P(\boldsymbol{s}_t = \boldsymbol{s}_t^{wt} | \boldsymbol{s}^{mt})] \qquad (4)$$

Here $T$ represents all mutants in $\boldsymbol{s}$ and $\boldsymbol{s}_t$ is the residue at position $t$ of the mutant and wild-type sequence. In this way, the initial performance of the supervised model is equal to its zero-shot performance, which makes it easier to fit the training data.

In each training iteration, we randomly select $m$ subsets of size $n$ with replacement from the training data ($n$ is smaller than the training data size, and $m$ can be viewed as batch size) and then compute ListMLE loss on each subset. The average loss on the $m$ subsets is used to update the trainable parameters in PLMs via gradient descent. The values of $m$ and $n$ depend on the actual training size and the validating performance on the training data.

**Meta-learning on auxiliary tasks**

Meta-learning aims to train a model that can quickly adapt to a new task using only a few data points and training iterations, typically by accumulating experience from learning multiple tasks[26,32,33]. Before transferring PLMs to the target few-shot learning task, we perform meta-learning to obtain a better initialization of the LoRA parameters that can learn the target training data faster and further reduce the possibility to overfit. To build the training tasks required for meta-learning, we search for existing labeled mutant datasets that are potentially helpful to predict the variant effects on the target protein, as well as generate pseudo labels through MSA for the candidate mutants.

**Searching similar experimental datasets.** Since ProteinGym has already included a wide variety of mutant data collected from 87 DMS assays, we use it as a database to search from. First, the wild-type sequences (or structures in the case of SaProt) of the target protein and the ones in the database are sent to a PLM (the one to be trained) to obtain their corresponding embedding vectors. Specifically, the output vectors representing each residue from the last hidden layer are averaged to form the protein embedding vector. Then, the relevance between the target protein and a candidate protein is measured by the cosine similarity between their embeddings $\boldsymbol{p}$ and $\boldsymbol{q}$:

$$\text{sim}(\boldsymbol{p}, \boldsymbol{q}) = \frac{\boldsymbol{p} \cdot \boldsymbol{q}}{\|\boldsymbol{p}\|_2 \cdot \|\boldsymbol{q}\|_2} \tag{5}$$

Finally, we choose two proteins that have the highest similarities to the target protein, and take their corresponding DMS datasets as the labeled data for the first two tasks.

**Estimate mutational effects based on MSA.** MSA is proved to be useful in predicting the mutational effects[35-38], and may also boost the performance of PLMs on various downstream tasks[11,19,31]. In this work, we integrate MSA knowledge into PLMs by meta-train PLMs on the pseudo labels generated by an alignment-based method, rather than modifying the model architecture. We choose GEMME algorithm[35] to utilize MSA information, which predicts mutational effects by explicitly modeling interdependencies between all positions in a sequence. It uses MSA to construct Joint Evolutionary Trees (JET)[48], and compute conservation degrees of different positions in the protein sequence based on evolutionary traces. The conservation degrees are then used to estimate the evolutionary fit required to accommodate mutations. The relative frequency of the mutation occurrence serves as another quantity to infer the mutational effect. Utilizing GEMME, we score the candidate mutated sequences of the target protein and build the dataset of the third task.

**Meta-training of PLMs.** We apply MAML[33], a state-of-the-art meta-learning algorithm, to enable PLMs to better utilize the few-shot training data. In essence, MAML learns to find the optimal initial model parameters such that small changes in them will produce large improvements on any target task, when altered by the target loss. Formally, we denote the fixed parameters of the original PLM and the added LoRA parameters as $\omega$ and $\theta$. Before training, for each task $\mathcal{T}_i$ we built, we randomly select half of its data as training data $\mathcal{D}_i^{tr}$ and the other half as testing data $\mathcal{D}_i^{te}$. Meta-training can be viewed as a bilevel optimization problem. In the inner-level optimization, a task $\mathcal{T}_i$ is randomly picked, and the loss on $\mathcal{D}_i^{tr}$ is used to temporarily update the current trainable parameters $\theta$ into task-specific parameters $\theta_i'$ by gradient descent. Assuming there is only one gradient update step, it can be expressed as:

$$\theta_i' = \theta - \alpha \nabla_\theta \mathcal{L}_{\mathcal{D}_i^{tr}}(f_{\omega,\theta}) \tag{6}$$

Here the loss function $\mathcal{L}$ and scoring function $f$ are the same as the ones in (2), and $\alpha$ is the step size. For the outer-level optimization, the parameters $\theta$ are trained by optimizing for the test performance of $f_{\omega,\theta_i'}$ across the tasks in current meta-batch:

$$\theta \leftarrow \theta - \beta \nabla_\theta \sum_i \mathcal{L}_{\mathcal{D}_i^{te}}(f_{\omega,\theta_i'}) \tag{7}$$

where $\beta$ is the meta step size. The above two levels of updates are performed repeatedly during the meta-training process until the stop criterion is satisfied. Through meta-learning, we aim to train a meta-learner $f_{\omega,\theta}$ that learns to learn similar tasks. In our experiments, we take 5 gradient steps for each inner loop, and each meta-batch includes one randomly sampled task. Since the outer update of (7) involves a gradient through a gradient, we use a first-order approximation introduced by Finn et al.[33] to reduce the computational cost. Stochastic gradient descent (SGD) is applied to solve the inner optimization and we use Adam[49] with default hyperparameters for the outer optimization.

## Foundation models

**ESM1v.** ESM-1v[12] is a transformer language model for variant effects prediction, which employs the ESM-1b architecture and masked language modeling approach of Rives et al.[41] It is trained on unlabeled protein sequences from Uniref90[50], with 650M parameters. It includes five models trained with different seeds for ensemble, but we only use the first checkpoint in our experiments.

**ESM-2.** ESM-2[14] introduces improvements to ESM-1b in architecture and increases the data for pre-training. In detail, it equips Rotary Position Embedding (RoPE)[51], and uses UniRef50 for training. We take the 650M version of ESM-2 for our experiments.

**SaProt.** Su et al. propose SaProt[16], a PLM trained with both protein sequence and structure data. They introduce a structure-aware vocabulary that combines both residue types and 3Di structure tokens encoded by Foldseek [52]. The pre-training dataset of SaProt consists of around 40M structures predicted by AlphaFold2[53]. SaProt employs the same model architecture as ESM-2, with the structure-aware tokens as input. In this work, we choose the SaProt version that is continuously pre-trained on PDB structures. For SaProt, the scoring function (4) is modified as follows so that it can fit the new vocabulary:

$$f(\boldsymbol{s}) = \frac{1}{|\mathcal{U}|} \sum_{t \in T} \sum_{u \in \mathcal{U}} [\log P(\boldsymbol{s}_t = \boldsymbol{s}_t^{mt} u | \boldsymbol{s}^{mt}) - \log P(\boldsymbol{s}_t = \boldsymbol{s}_t^{wt} u | \boldsymbol{s}^{mt})] \quad (8)$$

where $u$ is a 3Di token, $\mathcal{U}$ is the structure alphabet used by Foldseek, and $\boldsymbol{s}_t u$ is a structure-aware token in the new vocabulary.

## Datasets and evaluation

**ProteinGym benchmark.** ProteinGym is an extensive set of DMS assays for comparisons of different mutational effect predictors[31]. We evaluate the model performance on its substitution benchmark, which consists of about 1.5M missense variants from 87 DMS assays. Since the maximum input length of ESM-1v is 1024 tokens, we truncate the proteins which have more than 1024 amino acids and ensure that most of the mutations in their corresponding datasets occur in the resulting interval. To build the inputs for SaProt, we obtain the structures of the proteins via AlphaFold2 or download from AlphaFoldDB if available.

For each dataset in ProteinGym, we first randomly sample 20 single-site mutants as an initial training set. On the basis of these training examples, we add another 20 randomly selected single-site mutants to build the second training set. Analogously, we expand the training set size to 60, 80, and 100 for separate experiments. For each training set, all the remaining data (or part of them if we specify) are used as test set and we do not access them for hyperparameter selection. The above process for data splitting is repeated 5 times with different random seeds, and we report the average model performance for each training data size. Since the test data changes across the experiments, the performance of the zero-shot results is not constants.

**Early stopping by cross-validation.** Early stopping is widely used in deep learning to prevent a model from overfitting. When sufficient labeled data are available, it is generally based on a separate validation set. However, a held-out validation set may result in insufficient training data in a low-resource scenario. On the other hand, if the validation data size is assigned too small, the validation scores such as Spearman correlation may not be representative enough for early stopping. Based on these considerations, we propose to estimate the number of training iterations for transfer learning by Monte Carlo cross-validation[54]. Specifically, we create 5 random splits of the training set into training and validating data. The proportion of training and validating data is 0.5:0.5 when the training data size is less than 50 otherwise 0.75:0.25. For each split, the model is trained on the sub-sampled training data for up to

500 steps, and we record the Spearman correlation calculated on the validating data every 5 steps. After 5 rounds of training and validating, we choose the training step number with the highest average validation score across different splits, and finally train the model for that number of steps on the whole training data.

The training data from the target protein is also used to early stop the meta-training procedure. Similarly, 5 random splits of it are generated first. For every 5 steps of the outer optimization during meta-training, we train the current meta-learned model $f_{\omega,\theta}$ on the sub-sampled target training sets for 5 gradient steps (same to the inner optimization) and compute the validation Spearman scores. We stop meta-training if the average validation score of different splits does not improve within 20 consecutive records and pick the best meta-learner according to this score.

**Data availability**

The datasets used for experiments are from ProteinGym[31].

**Author contributions**

P.T. and L.H. designed and supervised this project. Z.Z and P.T. developed this method. Z.Z, L.Z and Y.Y wrote the code. Z.Z, L.Z., Y.Y., and M.L did the benchmarks. All of the authors reviewed and accepted the manuscript. Z.Z, L.Z, and Y.Y contributed equally to this work.

**Corresponding authors**

Correspondence to Liang Hong or Pan Tan.

**Funding**

This work was supported by the grants from the National Science Foundation of China (grant number 12104295, 11974239), the Innovation Program of Shanghai Municipal Education Commission (2019-01-07-00-02-E00076), Shanghai Jiao Tong University Scientific and Technological Innovation Funds (21X010200843), the Student Innovation Center at Shanghai Jiao Tong University, and Shanghai Artificial Intelligence Laboratory.

**Competing interests**

The authors declare no competing interests